\documentclass[conference]{IEEEtran}
\IEEEoverridecommandlockouts
\usepackage{cite}
\usepackage{amsmath,amssymb,amsfonts}
\usepackage{algorithmic}
\usepackage{graphicx}
\usepackage{textcomp}
\usepackage{xcolor}
\usepackage{listings}
\usepackage{multirow}
\usepackage{threeparttable}
\definecolor{lightforestgreen}{rgb}{0.13, 0.55, 0.13} 
\definecolor{lightergray}{rgb}{0.9, 0.9, 0.9} %

\lstdefinestyle{mystyle}{
    backgroundcolor=\color{lightergray},   
    basicstyle=\ttfamily\footnotesize\color{lightforestgreen},  
    frame=none,                      
    breaklines=true,
}
\def\BibTeX{{\rm B\kern-.05em{\sc i\kern-.025em b}\kern-.08em
    T\kern-.1667em\lower.7ex\hbox{E}\kern-.125emX}}
    
\begin{document}

\title{LLM-Based Identification of Infostealer Infection Vectors from Screenshots: The Case of Aurora}

\author{
    \IEEEauthorblockN{Estelle Ruellan}
    \IEEEauthorblockA{\textit{Flare Systems} \\
    Montreal, Canada}
    \and
    \IEEEauthorblockN{Eric Clay}
    \IEEEauthorblockA{\textit{Flare Systems} \\
    Montreal, Canada}
    \and
    \IEEEauthorblockN{Nicholas Ascoli}
    \IEEEauthorblockA{\textit{Flare Systems} \\
    Montreal, Canada}
}

\maketitle

\begin{abstract}

Infostealers exfiltrate credentials, session cookies, and sensitive data from infected systems. With over 29 million stealer logs reported in 2024, manual analysis and mitigation at scale are virtually unfeasible/unpractical. While most research focuses on proactive malware detection, a significant gap remains in leveraging reactive analysis of stealer logs and their associated artifacts. Specifically, infection artifacts such as screenshots, image captured at the point of compromise, are largely overlooked by the current literature. This paper introduces a novel approach leveraging Large Language Models (LLMs), more specifically gpt-4o-mini, to analyze infection screenshots to extract potential Indicators of Compromise (IoCs), map infection vectors, and track campaigns. Focusing on the Aurora infostealer, we demonstrate how LLMs can process screenshots to identify infection vectors, such as malicious URLs, installer files, and exploited software themes. Our method extracted 337 actionable URLs and 246 relevant files from 1000 screenshots, revealing key malware distribution methods and social engineering tactics. By correlating extracted filenames, URLs, and infection themes, we identified three distinct malware campaigns, demonstrating the potential of LLM-driven analysis for uncovering infection workflows and enhancing threat intelligence. By shifting malware analysis from traditional log-based detection methods to a reactive, artifact-driven approach that leverages infection screenshots, this research presents a scalable method for identifying infection vectors and enabling early intervention.

\end{abstract}

\begin{IEEEkeywords}
LLM, infostealer, malware
\end{IEEEkeywords}

\section{INTRODUCTION}
Infostealers are a type of malware that infect a victim computer, and steal all of the credentials, session cookies, and personal data out of a browser, in addition to other sensitive information from the host. As such, infostealer malware represents a major threat to corporate and personal identities today.

In 2024, Flare reported over 29 million (29,003,537) stealer logs posted on cybercrime forums and channels. The sheer volume of logs, each containing hundreds of credentials and multiple files per entry, renders manual analysis impractical. The overwhelming scale of data makes tracking and mitigating infostealer campaigns exponentially challenging for humans to do.

In recent years, many variants of infostealer malware have evolved beyond their data-exfiltration capabilities. A notable development is the inclusion of a screenshot-capturing functionality, which enables threat actors to take a snapshot of the victim's device. These screenshots are typically captured at or shortly after the point of infection, with the precise timing depending on the offset selected by the attacker.

\begin{table*}[h!]
\centering
\begin{tabular}{|l|c|c|c|c|}
\hline
\textbf{Malware Family} & \textbf{Nb Logs} & \textbf{Nb Logs with Screenshot} & \textbf{Nb Logs Non–Commercial Screenshot} & \textbf{\% Logs Non–Commercial Screenshot} \\
\hline
Redline        & 23,364,792 & 10,256,776 & 9,977,394  & 42.70\% \\
Unknown        & 14,083,695 & 1,663,249  & 1,621,479  & 11.51\% \\
Vidar          & 6,858,716  & 2,746,922  & 2,706,702  & 39.46\% \\
Lummac2        & 7,181,395  & 227,680    & 116,358    & 1.62\% \\
Raccoon        & 3,212,467  & 378,309    & 378,301    & 11.77\% \\
Stealc         & 3,789,336  & 382,817    & 322,349    & 8.51\% \\
Risepro        & 2,609,151  & 572,709    & 546,668    & 20.94\% \\
Nexus          & 843,693    & 77,451     & 61,813     & 7.33\% \\
Bradmax        & 293,259    & 77,631     & 73,817     & 25.17\% \\
Rhadamanthys   & 52,397     & 15,529     & 14,949     & 28.54\% \\
Aurora         & 125,079    & 78,271     & 77,916     & 62.33\% \\
\hline
\textbf{Total} & 62,414,980 & 16,477,344 & 15,835,846 & 25.37\% \\
\hline
\end{tabular}
\caption{Global Stealer Logs Statistics}
\label{tab:malware logs}
\end{table*}

Flare's dataset has amassed over 60 million stealer logs to date, capturing infections across millions of devices (see Table~\ref{tab:malware logs}). In particular, more than a quarter of these logs, approximately 16.5 million entries, include a "Screenshot" file.  In other terms, over 25\% of stealer logs contain a visual record of the crime scene at the moment of infection, providing comprehensive clues and evidence critical to understanding the infection. These screenshots have the potential to deliver immediate insights that can reveal context and subtleties often missed or overlooked in textual logs.

What may seem like a trivial feature from the perspective of the attacker—an intrusive snapshot of the victim's screen—has proven to be an unexpected gold mine for the cyber threat intelligence community. Initially, these screenshots may have served a simple purpose for threat actors: to gauge the effectiveness of their infection tactics and determine which traps were most successful. However, as malware campaigns have become increasingly numerous, this seemingly small addition has become a powerful tool for understanding and tracking infostealer campaigns. These screenshots offer unfiltered insights into the victim’s environment at the moment of infection. They can reveal critical information such as the webpage visited by the victim when the infection occurred or even the installer of a software, providing invaluable context leading to the infection. 

These untapped "crime scene" images represent a valuable resource for further analysis and investigation. They offer a unique visual metric that helps analysts identify and understand the infection vectors responsible for compromising millions of devices worldwide. Far from being a mere byproduct of the attack, these screenshots now represent a key source of intelligence for mapping, analyzing, and better mititgate infostealer campaigns.

\section{PREVIOUS WORK}
The continuous evolution of malware analysis has given rise to diverse detection methodologies over the past decades. Malware detection approaches can be broadly categorized into static signature-based and behavior-based methods (dynamic or memory analysis) ~\cite{nataraj_malware_2011, chakravarty_study_nodate, chen_understanding_2019, maniriho_systematic_2024}.

Signature-based detection remains a fundamental approach, where binary patterns extracted from malicious files serve as identifiable fingerprints.  While efficient for known threats, this traditional method relies on matching suspicious files against an established database of malware signatures. And so, signature-based detection is efficient for known malware but struggles with unknown threats ~\cite{nataraj_malware_2011, noauthor_comprehensive_nodate,maniriho_systematic_2024}.

On the other hand, behavior-based approaches use behavioural activities and patterns, like deleting, creating, or altering systems file operations, to detect malware. Behaviour-based detection represents a more dynamic approach to malware identification, focusing on runtime program activities rather than static signatures ~\cite{maniriho_systematic_2024}. By flagging deviations from normal operational patterns, this approach can detect both known and zero-day malware attacks. Malware behaviour can be captured through dynamic or memory analysis techniques ~\cite{maniriho_systematic_2024}. While static analysis struggles with obfuscated or modified malware that lacks known signatures, behavior-based analysis, even if more effective against packed or encrypted malware, is slower and may miss detections if certain conditions aren't met ~\cite{nataraj_malware_2011, chen_understanding_2019}. 

A vision-based approach to malware analysis emerged in 2011, when Nataraj and colleagues (2011) introduced a pioneering classification method that converts malware binaries into grayscale images. This approach leverages the observation that binaries from the same malware family, when visualized as images with pixel values between 0-255, exhibit consistent structural and textural patterns while differing notably across families. This computer-vision based approach provides a unique perspective as it requires neither disassembly nor code execution for classification. 

With the development of Large Language Models (LLMs), new methods have tried to use LLMs for malware detection and analysis. LLMs offer enhanced pattern recognition and contextual understanding, making them suitable for detecting evolving cyber threats ~\cite{rondanini_large_2024}. 

Al-Karaki and colleagues (2024) employed multiple LLMs for various applications including honeypots, text-based threats, code analysis, trend analysis, and disguised malware detection ~\cite{al-karaki_exploring_2024}. Building on this approach, Omar and colleagues (2024) extended the use of multiple LLMs to IoT malware detection, using network traffic analysis and anomaly identification ~\cite{omar_harnessing_2024}.

Sayed and colleagues (2024) focused on using LLMs to detect Domain Generation Algorithm and DNS exfiltration attacks~\cite{sayed_fine-tuning_2024}. Meanwhile, Sánchez and colleagues (2024) coupled LLMs with transfer learning to detect system call-based malware, achieving an accuracy and F1 score of approximately 0.86~\cite{sanchez_sanchez_transfer_2024}. Zahan and colleagues (2024) demonstrated the effectiveness of LLMs in identifying malicious npm packages, achieving a 99\% precision rate~\cite{zahan_leveraging_2024}. Hossain and colleagues (2024) leveraged the Mixtral infrastructure to detect malicious Java source code, significantly outperforming traditional static analysis tools~\cite{hossain_malicious_2024}. 

Rondanini and colleagues (2024) explored the potential of LLMs in edge computing environments, addressing computational challenges in malware detection~\cite{rondanini_large_2024}. Finally, Zhao and colleagues (2025) introduced AppPoet, a multi-view LLM-assisted system for Android malware detection that generates semantic function descriptions and employs a deep neural network classifier to produce human-readable diagnostic reports, achieving 97.15\% accuracy and 97.21\% F1 score~\cite{zhao_apppoet_2025}.

Recent research has demonstrated the potential of LLMs in malware detection across diverse domains. Applications span various domains such as classifying malicious code ~\cite{hossain_malicious_2024}, identifying insecure mobile applications ~\cite{zhao_apppoet_2025}, IoT malware ~\cite{omar_harnessing_2024}, analyzing network traffic ~\cite{omar_harnessing_2024}, system call-based malware detection ~\cite{sanchez_sanchez_transfer_2024}, and package malware ~\cite{zahan_leveraging_2024}. 

While specific performance metrics varied, all studies reported positive results for LLM-based malware detection approaches. Several studies noted that LLM-based methods even outperformed traditional techniques~\cite{hossain_malicious_2024} or achieved high accuracy scores~\cite{sanchez_sanchez_transfer_2024, zahan_leveraging_2024, zhao_apppoet_2025}.

Although extensive research has focused on proactive malware detection at the device level, malware infections continue to proliferate at an unprecedented pace. Current literature on malware detection has largely ignored a crucial reactive approach, analyzing victim-sourced artifacts like stealer logs and screenshots, to track malware campaigns. This gap reveals an overemphasis on proactive methods while neglecting valuable insights from compromised systems.
Moreover, despite recent advances in LLM-driven malware detection, no studies have explored their use for in-context analysis of these artifacts, presenting a novel opportunity for enhanced threat intelligence and detection.

\section{PRESENT RESEARCH}
 
Analyzing stealer logs presents significant challenges due to the overwhelming volume of credential data. Stealer logs can contain several hundreds of compromised credentials, making manual analysis impractical, especially given the sheer volume of logs available. Identifying infection vectors and tracking successful malware campaigns at scale remains a significant challenge.

Recognizing the limitations of traditional prevention and detection methods, our research pivots to leveraging artifacts of infection, specifically, stealer logs' screenshots. These screenshots represent the "crime scene" of digital infections, containing visual evidence of compromise. This white paper introduces a novel approach using a LLM to process screenshots of infected devices, typically captured at the point of infection, enabling efficient identification  of IoCs, infection vectors, and infostealer campaign tracking. By analyzing these post-infection artifacts using a LLM, we aim to identify and mitigate the infection vectors responsible for widespread device compromises, transforming victim screenshots from passive  artifacts into active threat intelligence tools, addressing a critical gap in current cybersecurity research. The primary objectives are as follows:

\begin{enumerate}
    \item \textbf{Performance Evaluation}: 
Measure the accuracy and reliability of the LLM in analyzing malware-related screenshots.
\item \textbf{Aurora Infostealer Analysis}: Investigate the Aurora infostealer infection campaign(s) by leveraging its infection screenshots to uncover key insights.
\end{enumerate}

By shifting the focus from traditional log-based analyses to the direct evaluation of infection screenshots, this method enables faster processing and more efficient identification of infection patterns, ultimately enhancing threat intelligence and incident response capabilities.

This paper focuses on infection screenshots from devices compromised by the Aurora infostealer. This focus is motivated by the observation that over 60\% of logs include an infection snapshot (see Table~\ref{tab:malware logs}). Consequently, analyzing the Aurora infostealer screenshots offers the potential to capture a more comprehensive view of infection campaigns and identify a broader range of infection vectors and IoCs compared to malware families where less than 10\% of logs contain screenshots.

\section{METHODOLOGY}
\subsection{Data Collection}
Infection screenshots were obtained through Flare’s Platform. Flare is an information technology (IT) security company that maintains a cyber threat intelligence platform by monitoring various online spaces\footnote{https://flare.io/}. A total of 1,000 screenshots from Aurora stealer logs were initially collected. 

\subsection{Analysis Framework}

After thorough screening, the images were encoded in base64 format before being processed by a LLM. Specifically, the gpt-4o-mini model was used to analyze the screenshots.

Our initial screening process provided valuable insights into the types of screenshots we would need to analyze. Through this preliminary examination, we identified three distinct categories of screenshots:

\begin{enumerate}
    \item Web Content Screenshots: Images primarily showing browser windows with open tabs, URLs, and web page content
    \item File System Screenshots: Images displaying file explorers, installers, or download interfaces
    \item Hybrid Screenshots: Images containing both web elements and file system components
\end{enumerate}

This categorization informed our prompt engineering strategy for the LLM. We specifically designed prompts to identify and extract two critical types of information: Web-related identifiers (browser tabs, URLs, domain names, and web page titles) and file system elements (installer executables, archive files, download queues, and file explorer contents).

By targeting both web and file system elements in our prompts, we ensured comprehensive coverage across all three screenshot categories. This dual-focus approach maximized our ability to extract relevant information for infection vector analysis. The model was provided with the following prompt: 
\begin{lstlisting}[style=mystyle]

"""
The images provided are screenshots of computer screens when they were infected by infostealer malware. Describe what is on the screen following this format:

    ### Main Content:
        Describe the main content visible on the screen, include as much detail as possible.

    ### Files/Programs:
        Installer: Focus on installers or install window, put the name of the file being installed. When there is a name for the installer window, get the name of file/folder or the path. 
        File explorer: Focus on file explorer if there is one. Put the names of files and their extensions in this section. If the path of the file explorer reveals the name of a folder/file, get it.
        Ignore all desktop programs and icons. Seperate filenames by a ",". If there aren't any file, executable or progam put "X".

    ### URL
        Put all URLs you see. If there aren't any URLs, put "X".

    ### Browser Tabs Analysis:
        Ignore bookmarks. For each active browser tab in the top row, list in this format:
        - [logo: {logo name}] [text: {visible text}] (meaning/context if apparent). If there aren't any webpage, put "X".

    ### Suspicious Elements:
        Highlight any file, executable, program, URL or download link that could contain malware. These could be youtube videos, blogs, google drive, etc.

    ### Language and Date:
        - **LANGUAGE:** language
        - **DATE:** date 
        
"""
\end{lstlisting}

The model subsequently generated a detailed description of the screenshot following the format given to the model. The screenshot shown in Figure~\ref{fig: screen_1} was processed by the model, producing the following description:

\begin{figure}[t] 
\centering 
\includegraphics[width=\columnwidth]{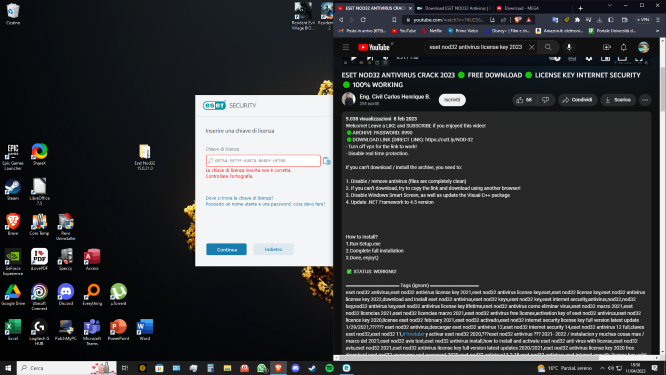}
\caption{Example of stealer log screenshot fed to the model.}
\label{fig: screen_1}
\end{figure}

\vspace{0.5cm} 

\begin{lstlisting}[style=mystyle]
    Content: ### Main Content:
The screenshot displays a computer desktop environment with an ESET Security window on the left side, prompting the user to input a license key. There is a red error message indicating that the entered license key is not correct. On the right side, there is a YouTube video titled "ESET NOD32 ANTIVIRUS CRACK 2023" which claims to offer a free download of a license key for ESET antivirus software. It includes a description with instructions for installation and cautionary steps such as disabling antivirus and Windows Smart Screen.

### Files/Programs:
Installer: ESET NOD32 ANTIVIRUS CRACK 2023  
File explorer: X

### URL
1. https://www.youtube.com/watch?v=HBG5nZQ7ThA  
2. Download ESET NOD32 ANTIVIRUS CRACK 2023  
3. License key internet security 100% working 

### Suspicious Elements:
- The YouTube video titled "ESET NOD32 ANTIVIRUS CRACK 2023" and the associated download link (https://cutt.ly/NOD-32) are highly suspicious, as they suggest accessing cracked software, which typically contains malware.
- The license key entry prompt in the ESET window may indicate that the user is following instructions from the video to illegally activate software.

### Language and Date:
- **LANGUAGE:** Italian
- **DATE:** 11/04/2023

\end{lstlisting}

\subsection{Assessment Framework}

The cornerstone of this research was evaluating the capability of the LLM to analyze infection screenshots for malware campaign tracking and attribution. Our primary objective was to assess whether an LLM could accurately describe infection screenshots and identify critical elements that enable campaign tracking and infection vector identification.

To assess the accuracy of the LLM system, we developed a comprehensive evaluation framework and applied it to a random sample of 100 screenshots from our dataset. The assessment focused on four key aspects:

\begin{enumerate}
    \item \textbf{General Scene Description:} The ability to understand and describe the overall context of the screenshot.
    \item \textbf{Browser Tab Identification:} The detection and analysis of open browser tabs and their content.
    \item \textbf{File Identification:} The recognition of files and system artifacts.
    \item \textbf{Suspicious Element Detection:} The identification of IoCs, infection vectors, and malicious patterns.
\end{enumerate}

The screenshots were fed to the LLM model discussed above. However, certain aspects did not apply to some screenshots (e.g. the LLM could not be assessed on Browsing Tabs Identification when the screenshot was just a windows desktop and a file explorer). To overcome this limit, we randomly selected additional screenshots to make sure that each aspect had at least 50 screenshots to be evaluated, bringing the assessment sample to 106 screenshots. 

\begin{table}[h]
\centering
\caption{Scoring System}
\begin{tabular}{|c|l|}
\hline
\textbf{Score} & \textbf{Description} \\ \hline
0   & Missing critical elements \\ \hline
1   &  Captured main elements but missed details \\ \hline
2  & Comprehensive capture of all relevant elements \\ \hline
99  & Not applicable to the specific screenshot \\ \hline
\end{tabular}
\label{tab:Scoring_system}
\end{table}

Then, two analysts went over each of the screenshots and assessed the accuracy of the description provided by the LLM. To ensure an objective assessment, the analysts followed a standardized scoring system; presented in Table~\ref{tab:Scoring_system}. They individually scored the four key aspects following the scoring system. Once individual coding was completed, the two analysts entered an agreement between the coders to obtain a common decision for each screenshot. Such intercoder agreement allowed them to pinpoint discrepancies, but also find areas of improvement to fine-tune the LLM, discussed below.

\subsection{Screenshot Description Analysis}
To analyze the screenshot descriptions, we performed a descriptive statistical analysis on URL and files/programs columns, to understand the distribution of values and identify emerging trends. For instance, we cross-referenced URLs with entries in the suspicious elements column to uncover overlapping instances. This analytical process allowed us to pinpoint frequently occurring patterns and potential IoCs, which were then subjected to further detailed investigation.

Campaign clustering methodology,
URL extraction and classification,
Campaign attribution techniques

\section{RESULTS}

\subsection{Performance of the LLM}

The LLM performed well in three out of four aspects. First, the LLM demonstrated exceptional accuracy (96\%) in providing \textbf{comprehensive scene descriptions} of the provided screenshots. The general descriptions of screenshots provided by the model demonstrated a strong contextual understanding of the scene and a high level of detail.

\begin{table}[htbp]
\centering
\caption{LLM Performance Assessment Results}
\label{tab:llm-assessment}
\begin{tabular}{|l|c|c|c|}
\hline
\textbf{Assessment Category} & \textbf{Score} & \textbf{Occurrences} & \textbf{Percentage} \\
\hline
\multirow{2}{*}{General Description} 
& 2 & 102 & 96.2\% \\
& 1 & 4 & 3.8\% \\
\hline
\multirow{4}{*}{Browser Tab Identification} 
& 2 & 15 & 14.1\% \\
& 1 & 16 & 15.1\% \\
& 0 & 18 & 16.9\% \\
& 99 & 55 & 51.9\% \\
\hline
\multirow{2}{*}{File Identification} 
& 2 & 90 & 84.9\% \\
& 99 & 16 & 15.1\% \\
\hline
\multirow{3}{*}{Suspicious Elements} 
& 2 & 90 & 84.9\% \\
& 1 & 12 & 11.3\% \\
& 0 & 2 & 1.9\% \\
& 99 & 2 & 1.9\% \\
\hline
\end{tabular}
\end{table}

Second, when files were present such as in a file explorer or downloaded files on a browser (85 applicable cases), the LLM achieved 100\% accuracy in comprehensive \textbf{file identification}. The LLM performed well in identifying file explorers of all types, even when partially obscured by other application windows. It was also able to recognize files downloaded from browser tabs.

Third, the LLM showed strong capability (87\%) in comprehensive \textbf{suspicious element detection}. Elements detected varied from URL, to video, to file or applications. The model was able to successfully identify suspicious elements and emit hypothesis as to why other elements could also be considered suspicious. It was also able to describe when nothing seemed suspicious as well, minimizing false positives.

However, the LLM struggled to recognize the nuance that files from the same date as the screenshot’s date—typically found on the desktop—are the most likely suspicious elements. In these cases, the LLM often flagged multiple files as potentially suspicious and was unable to pinpoint a single file to focus on. In contrast, a human would have been able to identify the source of infection based on the file's date.
\\\\
Finally, while the LLM proved highly efficient at identifying individual files and URLs, its performance in \textbf{browser tabs identification} was inconsistent. When browser tabs were present (50 applicable cases), all tabs were successfully identified in 30\% (15/50) of cases and partially identified in 32\% of cases (16/50). The LLM failed to accurately identify browser tabs in 36\% of cases (18/50). 

In some screenshots, the model perfectly identified tabs and accurately traced infection vectors, while in others it failed outright, misidentifying or overlooking critical information. This inconsistency in tab identification often impacted the model's ability to pinpoint the infection vector. Screenshots with complex browsing activity were occasionally analyzed with precision, but most of the time, the model's analysis was plainly incorrect.

The analysis of failure cases revealed specific patterns in the LLM's browser tab identification capabilities. Interestingly, in some cases, the model failed to identify all relevant tabs but successfully flagged text from a tab it had not recognized in the tab section as suspicious, despite not acknowledging it as a tab. Out of 34 cases where the LLM failed to accurately identify browser tabs, it still successfully identified the suspicious element in 26 instances. In 5 cases, the model was on the right track but missed crucial details, and in only 1 case, it completely failed to detect the suspicious element.  In the remaining 2 cases, even human reviewers were unable to identify any suspicious elements, and the screenshots were subsequently marked as '99' for suspicious elements. This finding underscores that, despite its inconsistency in identifying browser tabs, the model is highly effective at detecting suspicious elements.

Another notable challenge emerged in distinguishing between browser bookmarks and active tabs, particularly in screenshots featuring multiple open tabs. This limitation appears to be directly correlated with the visual constraints of modern browsers: as the number of open tabs increases, each tab's width decreases, resulting in truncated or completely hidden tab text. The reduced visibility of tab content forces the LLM to rely heavily on tab favicons (small website logos) for identification. However, this proves problematic for several reasons: favicon images are often too small for reliable recognition; many websites use generic or non-distinctive icons; and lastly, less popular websites may have unfamiliar logos that the model struggles to associate with specific content.

In scenarios with numerous open tabs, we observed that the LLM would consistently default to describing bookmark content instead of active tabs. This behavior is likely due to the bookmark bar maintaining full-text visibility regardless of the number of open tabs, making it a more accessible source of information for the model to process.


In certain cases, the model struggled to accurately determine the specific purpose of each tab or recognize logos, such as those representing Facebook or YouTube searches. Additionally, when multiple similar tabs were present (e.g. two separate tabs for Facebook), the model occasionally identified only one while overlooking the others. This limitation is problematic, as it results in the omission of potentially relevant information, which could hinder the accurate identification of infection vectors. 
\\

\subsection{LLM-generated Description Analysis}
After evaluating the LLM's performance, we proceeded with analyzing the LLM-generated screenshot descriptions. Leveraging the descriptions, we extracted 363 unique URLs, providing a substantial dataset of potential IoCs. After filtering out benign domains (e.g., google.com, walmart.ca) and accounting for 26 truncated URLs, we obtained 337 unique, actionable URLs for analysis. The extracted URLs fell into three main categories:

YouTube Videos (117 URLs):
While YouTube videos themselves are not inherently malicious, their description boxes often contain valuable potential IoCs.
These videos serve as intermediary vectors, potentially leading to malware distribution links. Which leads us to the second category of URLs.

File Distribution Platforms (65 URLs):
File hosting services (mega.nz, mediafire.com, anonfiles), shortened URLs (bit.ly, cutt.ly) and 
content delivery platforms (telegra.ph) were a recurring theme in screenshots. While most of them can be related to a youtube video, some of them were the only URL extracted of their respective screenshot. File distribution platforms URLs are the most actionable, as they likely provide direct access to the malicious file and should be prioritized for remediation.
Note: 6 URLs in this category were truncated

Other Domains (155 URLs):
Various platforms including GitHub repositories,
Cloud storage services (e.g., link.storjshare.io) and other diverse domains, potential phishing websites or webpages hosting the malware-laden archive directly.

The LLM-generated descriptions initially identified a total of 1,007 files across our screenshot dataset, consisting of 189 installer files and 818 other files. However, this raw extraction required further refinement to isolate potential indicators of malicious activity from benign files.

Not all files, aside from the installer files, were malicious or informative enough to pinpoint infection vectors. The LLM described both folders and every file visible in the file explorer, as per the instructions. In cases where a compressed archive was open on the screen, the LLM also listed all files within the archive. While archives often have descriptive names, the individual files inside typically have generic names, making them less useful for identifying infection vectors.

To establish a high-confidence dataset of potentially malicious files or informative files, we implemented a two-stage filtering approach. Initially, we cross-referenced all extracted file names with entries from the "Suspicious Element" section, flagging only those files that appeared in both datasets. This first pass allowed us to eliminate obvious non-malicious content. Subsequently, we conducted a manual review to exclude files with generic names that lacked distinctive characteristics (e.g., Setup.exe or win\_x64.dll).
This rigorous filtration process reduced the total file count from 1,007 to 246 relevant files. Within this refined dataset, we identified 239 unique files comprising: 79 executable (.exe) files, 38 compressed archives (.zip), 23 RAR archives (.rar), and 2 dynamic link libraries (.dll).

While finding these filenames is straightforward, connecting them to their source links, when missing from the screenshot, requires more work. This extra step is often needed to track down the harmful websites distributing these files. However, filenames alone can reveal significant information about the software and programs targeted by infostealer malware campaigns. While some filenames are too generic to determine the specific theme or infection method (e.g., Setup\_x64), others are highly descriptive (e.g., Microsoft\_Office\_Crack\_2022.rar), clearly indicating the lure used to entice victims. Notably, some filenames are so precise and detailed that a simple Google search can lead directly to distribution sites or related YouTube videos containing download links in their descriptions. This highlights how attackers likely leverage commonly searched software names to spread malware efficiently.

\subsection{Themes Used to Lure Users to Malware-laden Payloads}

Analyzing the descriptions as well as URLs and relevant files, we identified recurring lures, malware distribution methods and social engineering techniques used in successful infection campaigns. Through this approach, we uncovered three distinct and effective malware campaigns. By combining LLM-generated insights with qualitative analysis, we reconstructed the workflow of each campaign, detailing the social engineering tactics that facilitated infection. A comprehensive breakdown of these campaigns is provided in section~\ref{sec:notable_campaigns}. 

Our analysis revealed two primary themes driving device infections in our dataset: cracked software and gaming mods. These themes served as lures, enticing users to download malware under the guise of desirable content.

\subsubsection{Cracked Software}
The first and most prominent theme driving infostealer infection in our screenshots is cracked software. In our sample, 28.3\% of the infections were related to cracked software. In this theme, threat actors target users looking for free or cracked versions of often popular software or services, exploiting their desire to avoid paying for legitimate licenses. 

Software targeted are usually mainstream and widely recognized. Creative software such as Adobe Suite, Filmora, and VEGAS Pro, as well as AI-generated art platforms like Midjourney (with a detailed analysis of Midjourney provided in section~\ref{sec:midjourney}) are targets of choice for such schemes. However, the Office suite claims the top spot. Other software targeted included AnyDesk, Rufus, SketchUp and various versions of Office and Adobe Photoshop. 

Targeting mainstream software like the Adobe or Office suites ensures a wide pool of potential victims. Both suites have costly license, pushing some users to seek illegal alternatives. This behavior represents a lucrative opportunity for threat actors. Capitalizing on such behavior, threat actors disguise their malware as free, cracked versions of these widely-used software. By doing so, threat actors prey on users' willingness to bypass legitimate licensing fees at the cost of their own security, luring them into downloading malicious software disguised as cracked or free versions.

\subsubsection{Gaming Mods, Cheats, and Hacks}

The second theme groups infection vectors disguised as gaming mods, cheats and hacks. In this theme, malware is disguised as downloadable gaming mods or cheats, particularly for mainstream games to appeal to a larger pool of victims. Common examples include: \begin{itemize} 
    \item "Galaxy Skin Swapper" 
    \item "ModMenu PB.rar" 
    \item "Valorant AHK Triggerbot" 
    \item "Minecraft ModPack"
\end{itemize} 

Infections in our screenshots were related to gaming content in 7.4\% of cases. The choice to target mainstream games like Roblox, Minecraft and Fortnite is a strategic one. Since their release in 2006, 2009 and 2017 respectively, these games became a gateway into gaming for people of all ages. They are accessible, relatively simple, and their graphics are suitable even for younger audiences. The frequent introduction of paid skins  for characters and weapons, and unofficial mods to enhance gaming experience creates a strong appeal, especially among players who desire these features but lack the means or intention to pay for them. This demographic becomes an ideal target. Targeting mainstream games like Fortnite, Minecraft and Roblox is highly effective, as it ensures a large and susceptible audience.

Other elements found for this theme were related to games such as Valorant, Genshin Impact, Apex Legends, Minecraft, Grand Theft Auto V, FIFA 23, Terraria, Call of Duty, League of Legends, World of Tanks Blitz, Phasmophobia, ARK Survival Evolved, No Man's Sky, Sea of Thieves, Black Ops Cold War, Blox Fruits, and God of War, Sims 4. 

\subsection{Distribution Strategies}

\subsubsection{YouTube as a Distribution System}
\label{sec: youtube_distribution}
A recurring distribution strategy identified in our dataset is the use of YouTube videos, with 117 unique videos appearing in the screenshots. To be more specific, the links in the description box of those videos. Multiple screenshots display YouTube videos with titles such as:
\begin{itemize} 
    \item "BEST FORTNITE HACK UNDETECTED | FREE DOWNLOAD."
    \item "ADOBE AUDITION FULL CRACK 2022 FREE DOWNLOAD." 
\end{itemize} 

These YouTube videos take on the appearance of tutorials to guide users to malware-laden download links in their description, disguised as safe alternative options to download cracked software, gaming mods, hack or cheat. 

\begin{figure}[t] 
\centering 
\includegraphics[width=\columnwidth]{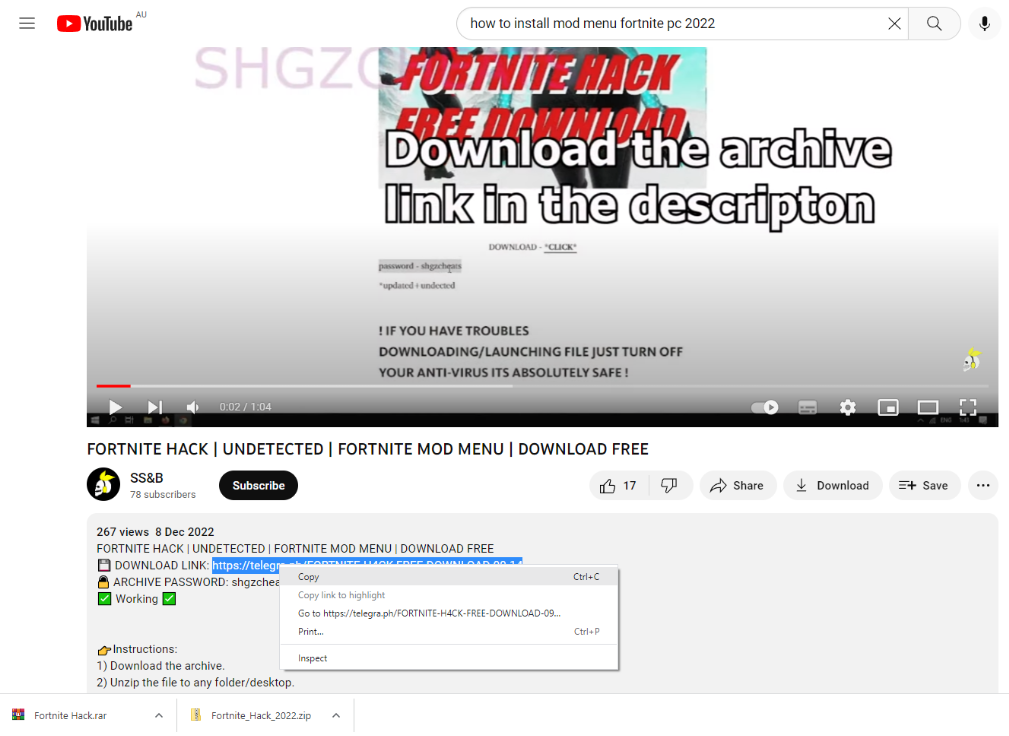}
\caption{Example of a YouTube video.}
\label{fig: screen_2}
\end{figure}

As shown in Figure~\ref{fig: screen_2}, a download link is usually available in the description, along with an archive password and instructions to follow. The video walks you through and demonstrates the process of installing the malicious payload, consistently emphasizing three key points:

\begin{itemize}
    \item The software is free 
    \item Disabling antivirus protection is necessary and "safe," for the installation 
    \item This cracked version/Game cheat works perfectly
\end{itemize}

YouTube is an ideal platform for distributing infostealer malware due to its massive global user base, with millions of users watching videos on a wide range of topics daily. Its audience spans all ages and interests, creating a vast pool of potential victims. Individuals often seek tutorials where someone walks them through every step of the process, and YouTube videos serve as the perfect medium for this. Threat actors exploit this demand for detailed visual instructions, preying on individuals searching for free or pirated versions of paid software and games mods. By embedding malicious links in video descriptions and guiding users through compromised downloads, they leverage YouTube’s credibility and accessibility to spread malware at scale.

\subsubsection{Leveraging Google Ads}

The last distribution strategy leveraged by the Aurora infection vectors is Google ad sponsoring of malicious websites. Built by threat actors, these web pages usually mimic the official version of the software or entity they are trying to impersonate. 

By sponsoring malicious sites for targeted keywords, they manipulate search results, positioning their fraudulent pages at the top. This technique capitalizes on users' inherent trust in top search results and official-looking websites, effectively tricking users into downloading malware under the disguise of legitimate software installations.

Used efficiently, Google Ads enable precise user targeting through sophisticated filtering mechanisms. Indeed, advertisers can specify the countries, languages, device types, demographics and even time of the day. This level of precision allows threat actors to maximize their reach and impact by tailoring their campaigns to specific user groups, ensuring malicious ads are shown to the most relevant pool of targets.

\subsection{Notable Campaigns}
\label{sec:notable_campaigns}
Leveraging the LLM-generated descriptions as well as qualitative analysis, we were able to piece together the steps of infection associated with three notable campaigns, namely: \textit{Blitz Java}, \textit{Zero MidJourney} and \textit{Snow Microsoft 2022}. 

\subsubsection{Blitz Java}

Java related IoCs were found in 5.7\% of infection screenshots. While most of them only featured a Java installer mimicking the official one, others revealed an interesting scheme. The Blitz Java campaign utilized social engineering techniques to trick users into downloading malware by masquerading as the legitimate and trusted Java installation.

The infected devices span the world, covering multiple languages, including English, French, Spanish, German, Swedish, Slovak, Portuguese, Polish, Hindi, Norwegian, Arabic, Korean, Dutch, Hungarian, and Thai. The Blitz Java playbook was the following: 

\begin{enumerate}
    \item Users search for Java download.
    \item Users click on a sponsored website that mimics the official Java platform.
    \item Users click the download button and receive a malware-laced Java\_Client.zip.
    \item Users execute the Java\_Setup executable
    \item The malware installs, stealing sensitive data
\end{enumerate}

The campaign's critical characteristic is its rapid timeframe, earning the name "Blitz" (German for lightning/flash). Infections occurred within a compressed window of 19 hours, specifically from February 11th, 2023, at 10:55 PM (CET) to February 12th, 2023, at 5:04 PM (CET).

The Blitz Java campaign is the perfect example on how to leverage Google Ads for malware distribution. The Blitz Campaign is based on malicious Google Ads for two domains impersonating Java's official download page: go.java-gapp.space and new.java-gapp.space. Both of which are no longer available (as of September 2024).

The ad campaign strategically targeted a weekend (February 11th and 12th falling on a Saturday and Sunday), which might appear coincidental but implies sophisticated targeting. Weekends typically offer individuals more leisure time, with reduced work and school commitments, increasing potential exposure and engagement.

Mimicking the legitimate Java.com site, the fake pages had subtle visual differences as shown in Figure~\ref{fig:}. The download page featured a button with an unusual file size. However, users wouldn't be aware of such differences unless they had visited the legitimate site beforehand. Clicking the "Download" button downloaded a "Java\_Client.zip" containing:

\begin{figure}[t] 
\centering 
\includegraphics[width=\columnwidth]{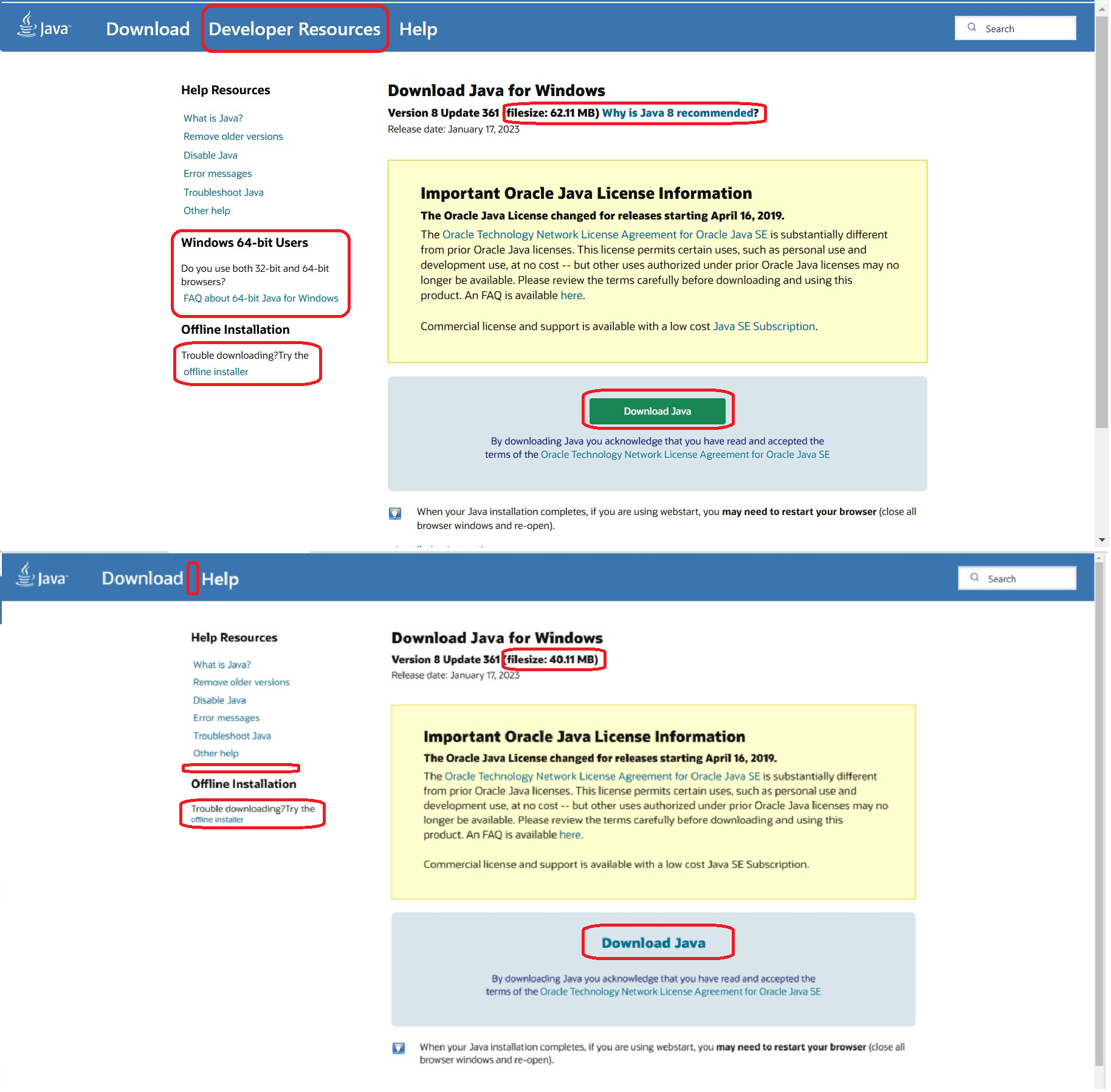}
\caption{Searching on the Web how to disable Antivirus}
\label{fig: javaVSjava.png}
\end{figure}

\begin{itemize}
    \item An executable (Java\_Setup.exe)
    \item A folder named “jre” with:
    \begin{itemize}
        \item A "bin" directory
        \item A "lib" directory
        \item Various license and readme files
    \end{itemize}
\end{itemize}

Once the executable was clicked, a Java installer appeared, triggering the installation of the malware. The installer closely replicated the legitimate Java installer, with one key difference: the taskbar logo was a black outlined cloud on a white background, unlike the official Java logo. By the time the installer appeared, the malware had already installed and successfully taken the infection screenshot.

Lastly, our analysis revealed that 18 out of 57 infected devices (31.6\%) downloaded Java for Minecraft-related purposes: specifically for Minecraft 1.19 crack, mod, shader pack, or performance enhancement via Optifine. Since Minecraft relies on Java for its core functionality (i.e. the game is originally developed using the Java programming language), this suggests a potential targeting of Minecraft players, though the hypothesis remains tentative.

\subsubsection{Zero MidJourney}
\label{sec:midjourney}
MidJourney artifacts were found in 6.3\% of our screenshots. Midjourney is an AI-based art generation platform, launched in July 2022 REFERENCE. Initially free to use, it has since transitioned to a subscription-based model. Surfing on this new wave of public interest, threat actors decided to capitalize on users’ desire to access Midjourney’s services for free. This campaign involved a scheme similar to the Blitz Java Campaign: setting up a typosquatting domain (ai.mid-j0urney.org), and advertising it, leveraging a likely malicious Google Ads account. 

The Zero MidJourney Campaign followed the playbook outlined below: 

\begin{enumerate}
    \item Users search for free access to Midjourney and click on a Google-sponsored ad for a fake domain (e.g., get.mid-journey.org).
    \item Users are redirected to a website mimicking Midjourney’s platform, offering a “Download for Windows” option.
    \item The page warns about antivirus alerts and advises users to disable their antivirus.
    \item Users download a malicious executable
    \item Users follow instructions to disable their antivirus.
    \item The malware installs, stealing sensitive data
\end{enumerate}

Our analysis identified several malicious domains mimicking Midjourney’s legitimate site, including “ai.mid-j0urney.org” and “get.mid-journey.com”. One of the primary infection vectors is a sponsored Google ad for get.mid-journey.org. The Google ad lured users into clicking the malicious domain despite user's browsers displaying a verified security badge for the legitimate site under the ad.

Once caught in the trap, users are presented with a web page mimicking Midjourney's official platform. The download page introduces a supposed “beta version” of the software, accompanied by a warning that the program may "falsely" trigger antivirus. The message emphasizes that AV triggers are "problems" common for all software in their beta version. 

This message is particularly interesting as it shows the conscious effort of threat actors to reassure users. By dismissing security alerts as "typical" and "expected," threat actors reassure users that disabling their antivirus software is a normal and necessary step for a "successful installation." This tactic significantly increases the likelihood of the malicious payload being installed successfully.

After downloading the executable and disabling their antivirus software, as advised, the installation proceeds. However, the software fails to run, instead prompting users to further disable their antivirus. Although this raises suspicion, users proceed since they were “warned” this would happen on the download page.

After following additional instructions to disable the antivirus, the system becomes vulnerable, and the infostealer malware executes. While users may initially hope for access to free Midjourney software, they soon discover that the program still doesn’t work. By then, their system has already been infected, leading to the theft of sensitive information such as passwords, cookies, and personal files.

Out of 1000 screenshots, 6.3\% (i.e. 63) involved the MidJourney campaign. Given that our data set was a random sampling of 1,000 screenshots from hundreds of thousands of weekly infections, and that malicious ads were being run against a well known and popular AI application, it is likely that this campaign was substantial and resulted in much more infections. 

\subsubsection{The Snow Microsoft 2022 Campaign}
\label{sec:snow_campaign}
While Microsoft Office cracks were found in 6.4\% of our screenshots, a recurring infection vector was especially related to a cracked version of Microsoft Office 2022: The Snow Microsoft Campaign.

The Snow Microsoft campaign had a wide pool of victims from all over the world. Screenshot from French, English, Spanish, Italian, German, Vietnamese, Japanese, Thai, Arabic, Polish, Dutch, Swedish, Turkish, Portuguese, Chinese and even Korean devices were identified. All these devices were infected by the same process: a cracked version of Microsoft Office 2022, laced with malware, provided in the description box of a single Youtube video. This campaign is a great example of Youtube as a Distribution system, as discussed in section~\ref{sec: youtube_distribution}

The choice to use a well-known, widely trusted, and essential software like Microsoft Office (which has a costly license, pushing some users to seek illegal alternatives) is a strategic one. Microsoft Office is recognized globally as the standard for basic computer tasks, with popular programs like Word, Excel, and PowerPoint included in the suite. Leveraging such a ubiquitous tool as bait to distribute infostealer malware provides access to an enormous pool of potential victims, spanning across the world. Even in countries with different alphabets, the word "Microsoft" is universally recognized. This makes it a perfect fit for one of the most potentially far-reaching malware campaigns to date. 

The Snow Microsoft Campaign follows the playbook outlined below: 

\begin{enumerate}
    \item Victim searches for “Microsoft Office 2022 crack” on YouTube.
    \item YouTube video from the DataStat channel lures victims with a download link for “free Office.”
    \item Link in the description box redirects to a Telegraph page promising free Office software.
    \item Malware hosted on MEGA.nz, users download a compressed archive.
    \item Archive is password-protected to avoid AV detection. Password is provided in the video's description box.
    \item Once extracted and executed, the malware installs infostealer to harvest sensitive information.
\end{enumerate}

First, users search for “Office [Year] crack” and click on the first video they are suggested: “Microsoft Office 2022 Crack \ Download Free\ Office 365 Free Version \ World Language” video from the DataStat Youtube Channel. DataStat Youtube channel is written in Azerbaijani and has 37 videos for a little over 2k subscribers (as of September 18th, 2024). The video is only a minute and 13 seconds long and presents a tutorial on how to download and install the cracked version of the software. However, the video is no longer available on the channel. 

In the video's description box, users are enticed with a download link claiming to offer a free version of the software. Clicking the link redirects users to a Telegraph page that promises a 'free version of Office.' The malware is hosted on MEGA.nz, a widely used file-sharing platform, with the provided link, leading to a compressed archive named 'Microsoft\_Office\_Crack\_2022.'

The archive consists of 
\begin{itemize}
    \item two Dynamic Link Library (DLL) files (win-32.dll and win-64.dll)
    \item An executable (@fomicvell.exe)
    \item a folder named “data”
\end{itemize}

The archive is protected by a password - "YUKI" - which is provided alongside the download link in the video's description box. This password protection is significant as it prevents anti-malware tools from scanning and detecting malicious content during the download process, proving effective in evading anti-virus detection. 

Finally, while downloading the infostealer malware, the user is presented with what looks like a legitimate Microsoft popup, enhancing the illusion of downloading legitimate software. Unbeknownst to them, they have installed an infostealer malware variant designed to harvest sensitive information.

\section*{Discussion}
This section examines three key results from this study: 1. the exploitation of users' trust as a persistent attack vector, 2. technical implementation challenges of LLM-based screenshot analysis, and 3. the effectiveness and resilience of our screenshot-based detection method.

\setcounter{subsubsection}{0} 
\subsubsection{Exploitation of User Trust: A Persistent Attack Vector}

Our analysis revealed an interesting pattern in Aurora's infection vectors: threat actors continue to rely predominantly on basic social engineering rather than sophisticated technical exploits. This finding underscores a fundamental vulnerability in the human element of cybersecurity. The campaigns we analyzed consistently exploited 1. users' willingness to compromise security for free versions of premium software, 2. trust in sponsored search results, and 3. reliance on video platforms like YouTube for software installation guidance.

The use of these simple yet effective tactics suggests that technical sophistication is often unnecessary when basic psychological tricks work so well. Particularly noteworthy were the Blitz Java and MidJourney campaigns, which demonstrated how sponsored search results can serve as a primary infection vector. Threat actors simply paid for Google ads, placing malicious content at the top of search results where users are most likely to click.

This reliance on mainstream platforms offers a strategic advantage for defenders. Campaign elements hosted on YouTube or delivered through Google ads can be reported and removed. These platforms have systems in place to remove harmful content once it's reported. While the Aurora screenshots in our dataset date back to 2020 (with most identified IoCs now inactive), this observation points to a promising defensive strategy:  if we can analyze screenshots from newly infected computers, we could enable early identification and takedown of active infection vectors, potentially limiting the spread of malware campaigns in their early stages.

\setcounter{subsubsection}{1}
\subsubsection{Technical Implementation Challenges}

A significant limitation emerged in the LLM's ability to fully analyze browser-based infection vectors. While the system demonstrated consistent performance in general scene description, files and URL identification, its tab analysis capabilities were inconsistent. This limitation has particular significance for threat intelligence gathering, as an important part of infection screenshots only feature tabs where much of the infection context is embedded in the combination and relationship between multiple browser tabs.

This finding suggests that optimal threat intelligence gathering from infection screenshots likely requires a hybrid approach: leveraging the LLM's strengths in rapid identification of files, URLs, and general scene description, while incorporating human analysis to address inconsistencies and decode complex browser-based infection chains where contextual understanding of tab relationships is essential. Feeding only the top 10\% of the image, which contains the tabs, separately to the LLM, followed by providing the entire screenshot, could offer a promising solution to address this challenge. This hybrid approach may enhance the LLM’s ability to analyze tab relationships more effectively, potentially improving the overall analysis of browser-based infection vectors.

Our prompt design prioritized comprehensive description and context, which inevitably generated substantial prose alongside the critical indicators. While effective for our initial analysis, we acknowledge this approach may not be optimal for high-volume processing. We encourage the research community to build upon our foundation and develop more targeted prompts that can extract specific potential IoCs more efficiently, particularly when seeking to identify a single file or URL per screenshot.

\setcounter{subsubsection}{2}
\subsubsection{Screenshot Analysis: Effectiveness and Resilience}
Despite limitations, the LLM-based analysis successfully streamlined the identification of several hundred potential IoCs. 

The screenshot-based approach embodies both our greatest strength and our primary limitation: its dependence on the existence and quality of the screenshot. As noted earlier, screenshots may be delayed, capturing irrelevant elements such as empty desktops or unrelated browser tabs. The timing and content of these captures significantly impact our analysis capabilities.

Lastly, ~\cite{chen_understanding_2019} highlights the continuous arms race between malware detection systems and emerging threats, emphasizing the need for algorithm robustness against adversarial modifications. While traditional detection methods must constantly adapt to code-level changes, our screenshot-based analysis offers inherent advantages in this regard. By focusing on visual artifacts rather than code signatures, our approach maintains effectiveness across diverse malware families. This visual analysis methodology is only vulnerable to a single point of failure: the complete removal of the screenshot functionality from the malware itself. As long as infostealers continue to capture screen content, our detection approach remains viable, even as the underlying code evolves.

\section*{Conclusion}

This research demonstrates the potential of leveraging LLMs to analyze infection screenshots as a novel approach to identifying malware infection vectors and tracking infostealer campaigns. By shifting the focus to infection screenshots, we introduce an alternative method that transforms passive forensic evidence into actionable threat intelligence.

 The analysis of Aurora infostealer campaigns confirms that simple social engineering techniques remains the primary cause of infection, with threat actors exploiting user trust in search engines, video platforms, and free software downloads. These mainstream platforms serve as both an attack surface and a potential defensive advantage
 for cybersecurity efforts, as malicious content can be reported and removed.

Key findings highlight strengths and challenges of this novel approach. While LLMs excel in extracting IoCs from screenshots, their performance in analyzing browser-based infections remains inconsistent. Addressing these limitations through hybrid methodologies could enhance 
the effectiveness of automated threat intelligence. Despite some limitations, our results underscore the resilience of visual-based detection. Unlike signature-based methods, which are susceptible to obfuscation
 and code changes, screenshot analysis remains viable as long as malware continues to capture screenshots. This approach offers a promising avenue for scaling malware campaign tracking and early intervention, enabling researchers, fraud, and security teams to rapidly identify new infostealer campaigns and distribution techniques.

\section{Acknowledgments}
The authors thank Nicholas Ascoli for originating the core idea that inspired this work, namely, the use of large language models to analyze screenshots extracted from infostealer logs. His initial insight laid the foundation for the project and guided its development toward the identification and of indicators of compromise at scale and tracking of successful campaigns.

\bibliographystyle{plain}
\bibliography{lit_review}

\end{document}